\documentclass[final,5p,times,twocolumn]{elsarticle}
\usepackage{amssymb}
\usepackage{amsmath}
\usepackage{multirow}
\biboptions{numbers,sort&compress}
\usepackage[hidelinks]{hyperref}
\usepackage{xcolor}
\usepackage{lipsum}
\usepackage{tabularx}
\usepackage{setspace}
\usepackage{booktabs, makecell}
\hypersetup{
	colorlinks,
	linkcolor={red!50!red},
	citecolor={blue!50!red}, %
	urlcolor={blue!80!red}
}
\usepackage{amsthm}
\usepackage{caption}

\journal{Physical Review Letters}
\makeatletter
\renewcommand{\theaffn}{\arabic{affn}}
\makeatother

\begin{document}

\begin{frontmatter}
\title{\textbf{Acoustic Analogy of Quantum Baldin Sum Rule for Optimal Causal Scattering}}

\author{Sichao Qu$^{1,*}$}
\author{Zixiong Yu$^{2}$}
\author{Erqian Dong$^{1}$}
\author{Min Yang$^{3,\dag}$}
\author{Nicholas X. Fang$^{1,4\ddag}$}

\affiliation[inst1]{organization={Department of Mechanical Engineering, The University of Hong Kong},
	addressline={Pokfulam Road}, 
	city={Hong Kong},
	country={China}}
	
\affiliation[inst2]{organization={Yau Mathematical Sciences Center, Department of Mathematical Sciences},
	addressline={Tsinghua University},
	city={Beijing},
	country={China}}
	
\affiliation[inst3]{organization={Acoustic Metamaterials Group Ltd.},
	addressline={Data Technology Hub, TKO Industrial Estate},
	city={Hong Kong},
	country={China}}

\affiliation[inst4]{organization={Materials Innovation Institute for Life Sciences and Energy (MILES)},
	addressline={HKU-SIRI},
	city={Shenzhen},
	country={China}}
\begin{abstract}
	\begin{center}
	\begin{minipage}{0.95\textwidth}
	\vspace{-2mm}
	The mass law is a cornerstone in predicting sound transmission loss, yet it neglects the constraints of causal dispersion. Current causality-based theories, such as the Rozanov limit, are applicable only to one-port reflective absorbers. Here, we derive a universal sum rule governing causal scattering in acoustic systems, establishing a rigorous analogy to the Baldin sum rule in quantum field theory. This relation reveals that the integral of the extinction cross-section is fundamentally locked by the scatterer’s static effective mass and stiffness, which is validated numerically using seminal examples of underwater metamaterials. Furthermore, the proposed sum rule predicts an optimal condition for an anomalously broadened transmission loss bandwidth, as experimentally observed through the spectral shaping effect of an acoustic Fano resonator. Our findings open up an unexplored avenue for enhancing the scattering bandwidth of passive metamaterials.
	\\
	{\small [Corresponding author(s): $^{*}$qusichao@hku.hk; $^{\dag}$min@metacoust.com; $^{\ddag}$nicxfang@hku.hk]}
	\end{minipage}
	\end{center}
\end{abstract}
\end{frontmatter}


\textit{Introduction.}— The transmission loss (TL) of a soundproofing wall with thickness \( L \) increases by 6 dB for each doubling of either the effective mass density per unit area (\( \rho_{\mathrm{eff}}L \)) or the angular frequency \( \omega \), known as the 'mass law': \( \mathrm{TL} \approx 20 \log_{10} \left( \frac{\omega \rho_{\text{eff}} L}{2 Z_0} \right) \), where the characteristic impedance of the background fluid is given by \( Z_0 = \sqrt{\rho_0 K_0} \) (with \( \rho_0 \) and \( K_0 \) representing the mass density and bulk modulus of the background fluid, respectively). In 2000, Liu et al. \cite{liu2000locally} challenged this principle by introducing lead spheres encased in a soft rubber and epoxy matrix, which exhibit deep-subwavelength dipole resonances characterized by dynamically negative mass density. This negative property \cite{lee2016origin} can lead to a bandgap with strong backscattering, even when the scatterer unit is significantly smaller than the excitation wave's wavelength, thereby violating the mass law. Fang et al. \cite{fang2006ultrasonic} further identified the monopole counterpart of negative bulk modulus through an engineered Helmholtz resonator array. In essence, metamaterials can concentrate modes within a specific band of interest, to create large range of dynamic properties, by leveraging the density of states from other frequencies \cite{monticone2013cloaked, wang2025seven,landi2018acoustic}. This enables exciting applications such as wave focusing \cite{oh2023engineering}, cloaking \cite{xu2021transformation}, imaging \cite{zhu2011holey}, tweezing \cite{xu2024acoustofluidic}, and perfect absorption \cite{qu2022microwave,yang2025acoustic}. However, the mass law disregards the causal dispersion of materials and could be inapplicable at excessively low frequencies (stiffness control region) or excessively high frequencies (where higher-order modes emerge). The well-known Rozanov limit addresses the causality-induced dispersion constraint on one-port absorption bandwidth \cite{rozanov2000ultimate, acher2009fundamental, yang2017sound}, yet it assumes a transmission-forbidden boundary. Thus, an open research question remains on the universal causal bound governing transmission loss: how the allocation of local-band spectral modes dictates the global wave transport?

\begin{figure}[t!]
	\centering
	\includegraphics[width=0.45\textwidth]{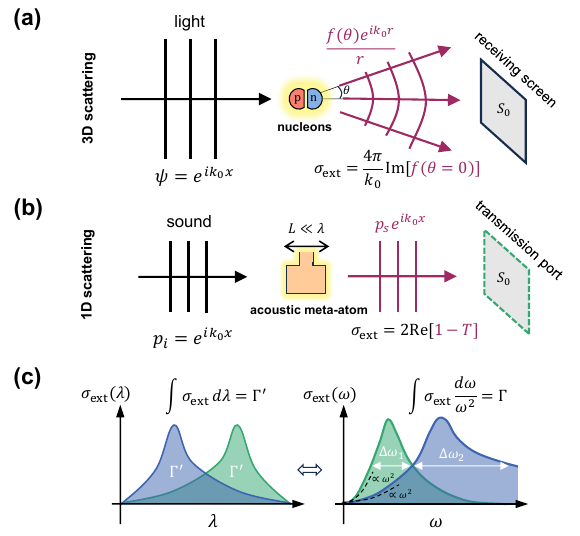}
	\caption{\label{fig:1} \color{black} From quantum to acoustic Baldin sum rule. (a) Schematic illustration of light–matter scattering. (b) Schematic illustration of sound–structure scattering. (c) Conceptual diagram of $\sigma_\mathrm{ext}$ spectra shaping, showing the conversion between frequency $\omega$ and wavelength $\lambda$.}
\end{figure}

To address this problem, we draw inspiration from micro-scale causal property of the quantum theory, the Baldin sum rule \cite{BALDIN1960310}, specifically a fundamental result in quantum field theory that connects the electric ($\alpha_e$) and magnetic ($\alpha_m$) polarizabilities of nucleons to the integral of the photo-absorption extinction cross-section $\sigma_\mathrm{ext}$: $\int_{\nu_0}^{\infty} \sigma_\mathrm{ext}  {d\nu/\nu^2} = 2 \pi^2(\alpha_e + \alpha_m)$ (natural unit),
where $\nu_0$ denotes the threshold energy to trigger nuclear light-matter interaction. As shown in Fig.~(\ref{fig:1}a), by measuring the far-field forward scattering amplitude $f(\theta=0)$, \textit{optical theorem} \cite{gustafsson2012optical}, a result of scattering unitarity, gives $\sigma_\mathrm{ext}=(4\pi/k_0)\mathrm{Im}[f(\theta=0)]$ ($k_0$ is the incident wavenumber). So, nucleon structure parameters can be directly extracted from scattering-based experimental data \cite{schumacher2005polarizability,holstein2014hadron}.

Here, we propose an acoustic analogy of quantum Baldin sum rule for for passive, linear, time-invariant sound-structure interaction:
\begin{equation} \label{eq:sum_rule}
	\int_0^{\infty} \sigma_\mathrm{ext}(\omega) \frac{d \omega}{\omega^2} = \Gamma,
\end{equation}
where $\sigma_\mathrm{ext}$ denotes extinction cross-section area (normalized by port area $S_0$), and $\Gamma$ contains acoustic effective properties (as the analogy of electric and magnetic polarizabilities of the quantum scatterer). The formulation of acoustic analogy bears a striking resemblance to the quantum Baldin sum rule [see Table~(\ref{tab:baldin_comparison}) in End Matter]. Next, we will derive Eq.~(\ref{eq:sum_rule}) and interpret its physical implications.

\textit{Model definitions.}— Acoustic metamaterials are commonly realized as periodic subwavelength structures or duct-embedded resonators for effective control of wave propagation. Accordingly, we focus on a one-dimensional (1D) scattering model [Fig.~(\ref{fig:1}b)], distinct from the three-dimensional (3D) spherical scattering scenario [Fig.~(\ref{fig:1}a)]. We define the complex transmission coefficient $T(\omega)$ as the ratio of the transmitted total field $p_t$ to the incident field $p_i$ across a structure of thickness $L$ (where $L \ll \lambda$), whose effective density and longitudinal modulus denote $\rho_{\mathrm{eff}}$ and $M_{\mathrm{eff}}$, respectively (see a homogenization method in Ref.~\cite{yang2014homogenization}). Thus, the normalized forward scattering amplitude is given by $T(\omega) - 1$, derived from the scattered field $p_s = p_t - p_i$, while $R(\omega)$ denotes the reflection coefficient or backward scattering amplitude. {\color{black}The extinction cross-section area is defined as the sum of the absorption term [$(1 - |R(\omega)|^2 - |T(\omega)|^2) S_0$] and the scattered backward and forward cross-sections [$(|R(\omega)|^2 + |T(\omega) - 1|^2) S_0$], where $S_0$ denotes the periodic wavefront/port area or duct cross-section area. Therefore, we reduce the extinction cross-section $\sigma_\mathrm{ext}$ (normalized by $S_0$) to}
\begin{equation} \label{eq:1Dot}
	\sigma_\mathrm{ext}(\omega) = 2\,\mathrm{Re}[1 - T(\omega)],
\end{equation}
which can be regarded as a 1D \textit{optical theorem} ({\color{black}derivations based on energy conservation are available in Ref.~\cite{labelSM}, Sec.~S1}). Since $T(\omega) - 1$ is analytic in the upper half-plane of frequency, we apply the Kramers–Kronig (KK) relations \cite{waters2000applicability} to obtain ($\mathcal{P}$ denotes the principal value)
\begin{subequations}\label{eq:kk}
	\begin{align}
		\mathrm{Re}[T(\omega)-1] &= \frac{2}{\pi} \mathcal{P} \int_{0}^{\infty} \frac{\omega^\prime \mathrm{Im}[T(\omega^\prime)-1]}{{\omega^{\prime}}^2-\omega^2} d \omega^{\prime}, \label{eq:kk_a} \\
		\mathrm{Im}[T(\omega)-1] &= \frac{-2\omega}{\pi} \mathcal{P} \int_{0}^{\infty} \frac{\mathrm{Re}[T(\omega^\prime)-1]}{{\omega^{\prime}}^2-\omega^2} d \omega^{\prime}. \label{eq:kk_b}
	\end{align}
\end{subequations}
Meanwhile, the analytical expression of $T(\omega)$ is \cite{labelSM}
\begin{equation} \label{eq:T}
	T(\omega) = \left[ \cos \left(k_{\mathrm{eff}} L\right) - \frac{i}{2} \left( \frac{Z_{\mathrm{eff}}}{Z_0} + \frac{Z_0}{Z_{\mathrm{eff}}} \right) \sin \left(k_{\mathrm{eff}} L\right) \right]^{-1},
\end{equation}
where the effective wavenumber $k_{\mathrm{eff}} =\omega/c_{\mathrm{eff}}=\omega\sqrt{\rho_{\mathrm{eff}}/M_{\mathrm{eff}}}$ and effective impedance $Z_{\mathrm{eff}}=\sqrt{\rho_{\mathrm{eff}}M_\mathrm{eff}}$. Note: if the assumptions holds: $Z_\mathrm{eff}\gg Z_0$ and $\omega \rho_{\text{eff}} L\gg Z_0$ ($\rho_{\text{eff}}$ is a real-valued constant), the transmission loss ($\mathrm{TL}=-10 \log _{10}|T(\omega)|^2$), can be approximated as $	 20 \log_{10} \left( \omega \rho_{\text{eff}} L/2 Z_0 \right)$ (mass law).

Because $\lim_{\omega \to 0} T(\omega) = 1$ for arbitrary scatterers with finite thickness and material properties, Eq.~(\ref{eq:kk}a) yields $\int_{0}^{\infty} \mathrm{Im}[T(\omega)-1] d \omega / \omega = 0$ (this is relatively trivial without the information of the scatterer). By inserting Eq.~(\ref{eq:1Dot}) into Eq.~(\ref{eq:kk_b}) and taking $\omega \to 0$, we derive acoustic Baldin sum rule in Eq.~(\ref{eq:sum_rule}), with the bound explicitly defined as
\begin{equation} \label{eq:bound}
	\Gamma = \pi \lim_{\omega \to 0} \frac{\mathrm{Im}[T(\omega) - 1]}{\omega} = \frac{\pi L}{2 c_0} \left(\frac{K_0}{M_{\mathrm{eff}}(0)} + \frac{\rho_{\mathrm{eff}}(0)}{\rho_0}\right) ,
\end{equation}
where sound speed $c_0 = \sqrt{K_0 / \rho_0}$. Defined at the static limit ($\omega \to 0$), $\Gamma$ can be split into a monopole term $\Gamma_m=({\pi L}/{2 c_0})  {K_0}/M_{\mathrm{eff}}(0)$ and a dipole term $\Gamma_d=({\pi L}/{2 c_0}) \rho_{\mathrm{eff}}(0)/{\rho_0} $. Acoustic Baldin sum rule reasonably eliminates the possibility of full-band perfect transmission suppression when $\Gamma\neq0$: if \( |T(\omega)| = 0 \) for arbitrary $\omega$, the integral in Eq.~(\ref{eq:sum_rule}) would diverge.

By taking \textit{subwavelength approximation} ($k_{\mathrm{eff}} L \ll 1$), the second-order Taylor series of Eq.~(\ref{eq:T}) is
\begin{align} \label{eq:T_low}
	T(\omega) = 1+ i\omega\frac{\Gamma_m+\Gamma_d}{\pi} -\omega^2 \frac{\Gamma_m^2+\Gamma_d^2 }{\pi^2} +i\mathcal{O}(\omega^3),
\end{align}
where the first-order coefficient is locked via $ \Gamma_m+\Gamma_d$ (sum rule's bound $\Gamma$), while the second-order coefficient determines the low-frequency asymptotic behavior: $\sigma_\mathrm{ext}(\omega)= \omega^2 [2 (\Gamma_m^2+\Gamma_d^2)/{\pi^2}] +\mathcal{O}(\omega^4)$, according to Eq.~(\ref{eq:1Dot}). 

We discover an opportunity for bandwidth enhancement, as shown in Fig.~(\ref{fig:1}c). The $\sigma_\mathrm{ext}$ integral ($\Gamma^\prime$) over the wavelength ($d\lambda \propto d\omega/\omega^2$)  remains the same for a scatterer before (green curve) and after (blue curve) optimal spectral shaping via the minimization of second-order coefficient $[2 (\Gamma_m^2+\Gamma_d^2)/{\pi^2}\geq4 \Gamma_m\Gamma_d/\pi^2]$. Due to the conservation of integral area of $\sigma_\mathrm{ext}/\omega^2$, the bandwidth in the right-side case can be much larger, i.e., $\Delta\omega_2 \gg \Delta\omega_1$.

\textit{Analytical validation.}— We investigated typical cases of ideal point scatterers, with closed-form dispersion models of $\rho_{\mathrm{eff}}(\omega)$ and $M_{\mathrm{eff}}(\omega)$. Mathematically, we have analytically obtained perfect agreement between $\int_0^{\infty} \sigma_\mathrm{ext}(\omega) \frac{d \omega}{\omega^2}$ and $\Gamma$ for all cases (see derivations in Ref.~\cite{labelSM}, Sec. S1).

\textit{Numerical validation.}— {\color{black}Ideal analytical models, as given in Eq.~(\ref{eq:T}), neglect higher-order modes (in waveguide scattering) or diffraction modes (in planar periodic samples). However, the sum rule does not preclude the existence of dynamic modes beyond the effective medium description, since all scattering states remain orthogonal.}

\begin{figure}[t!]
	\centering
	\includegraphics[width=0.48\textwidth]{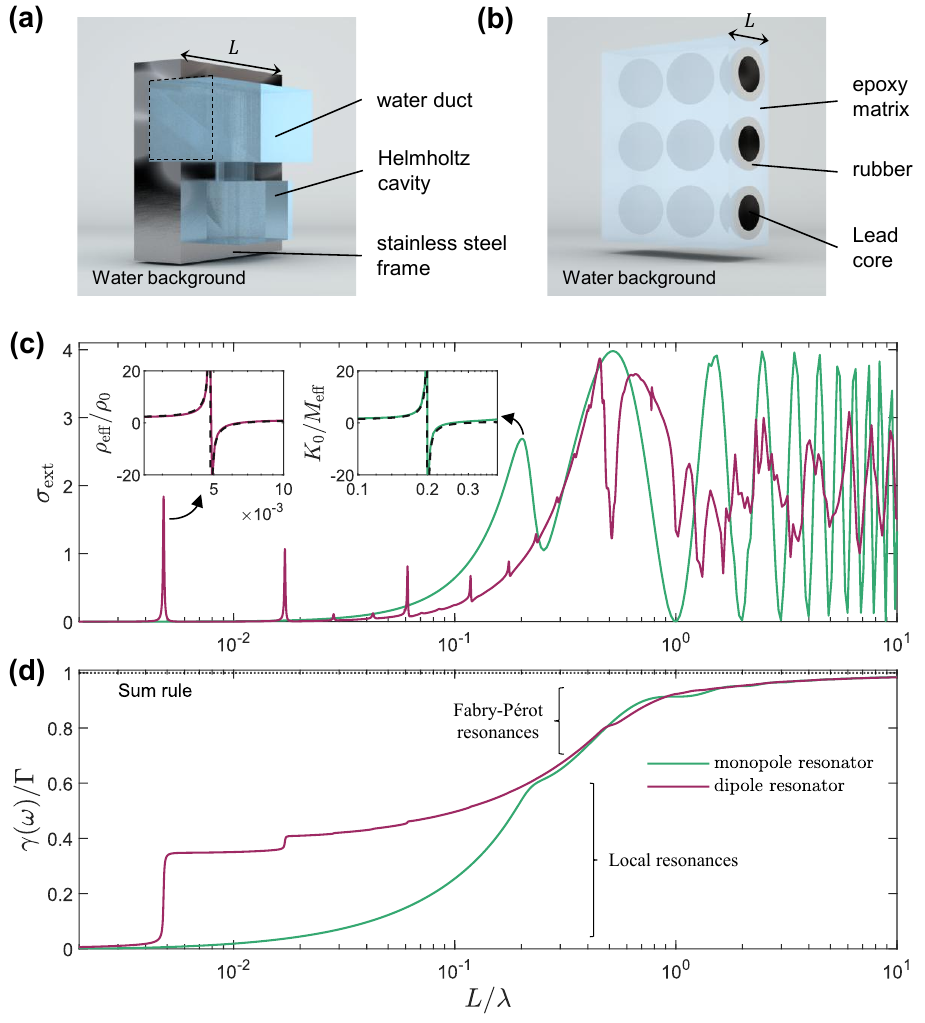}
	\caption{\label{fig:2} \color{black} The simulation-based verification of acoustic Baldin sum rule by revisiting seminal examples of underwater metamaterial scatterers. (a) The monopole Helmholtz resonator in a duct. (b) The dipole lead-core resonator in a periodic layout. (c) Extinction spectra ($\sigma_\mathrm{ext}$); insets show the extracted effective properties. (d) The cumulative distribution function $\gamma(\omega)$.}
\end{figure}

{\color{black} Here, the numerical examples we investigated include both a monopole Helmholtz resonator \cite{fang2006ultrasonic} and a lead-core dipole resonator \cite{liu2000locally} that incorporate higher-order fluid-based or elastic modes [see Fig.~(\ref{fig:2})].} Using material parameters from the original studies, we performed finite element simulations (FEM) to calculate $\sigma_\mathrm{ext}(\omega)$, with water set as the background media (see Ref.~\cite{labelSM}, Sec.~S2 for simulation details). As shown in Fig.~(\ref{fig:2}c), the monopole resonator—mounted on a duct—exhibits two subwavelength peaks: the first corresponds to the Helmholtz resonance ($L/\lambda \approx 1/5$), and the second to the half-wavelength Fabry–Pérot resonance ($L/\lambda \approx 1/2$) within the defined structural region of length $L$. For $L/\lambda > 1$, oscillations arise due to higher-order Fabry–Pérot resonances and associated phase modulation. Similarly, the dipole resonator array, shows multiple peaks ($L/\lambda$ as small as $1/200$), attributed to the elastic modal behavior of composite materials~\cite{still2013soft}.  After extracting $M_{\mathrm{eff}}(\omega)$ and   $\rho_{\mathrm{eff}}(\omega)$ spectra via $R(\omega)$ and $T(\omega)$ using the method in Ref.~\cite{groby2021analytical}, we observe that the corresponding lineshapes near resonance follow the Lorentz model, as shown in the insets of Fig.~(\ref{fig:2}c).

Next, to measure how fast the scattering resources can accumulate over frequency, we introduce the cumulative distribution function
\begin{equation}
	\gamma(\omega) = \int_0^{\omega} \sigma_\mathrm{ext}(\omega) \frac{d\omega}{\omega^2},
\end{equation}
which approaches the bound $\Gamma$ as $\omega \to \infty$ [$\gamma(\omega)/\Gamma \to 1$ for both cases in Fig.~(\ref{fig:2}d)]. We adopted the effective medium theory of composite materials \cite{zhou2009analytic, milton2022theory} to calculate $\Gamma$ with the static normalized effective density ${\rho_{\mathrm{eff}}(0)}/{\rho_0}$ and compressibility ${K_0}/{M_{\mathrm{eff}}(0)}$ [see Ref.~\cite{labelSM}, Sec. S2].  By comparing Figs.~(\ref{fig:2}c) and (\ref{fig:2}d), we can clearly see the local-band contribution in $\gamma(\omega)$ from the low-frequency local resonances to the higher-order Fabry–Pérot resonances.

\textit{Design strategy.}— The acoustic Baldin sum rule serves as a predictive tool: by maximally suppressing the low-frequency $\sigma_\mathrm{ext}$, scattering resources are redistributed to higher frequencies, broadening the operational bandwidth [Fig.~(\ref{fig:1}c)]. To quantify this spectral shaping effect, we combine Eq.~(\ref{eq:T_low}) and Eq.~(\ref{eq:1Dot}) and then compute $d\gamma(\omega)/d\omega$ ($={\sigma_\mathrm{ext}(\omega)}/{\omega^2} $) at the low frequency limit
\begin{align} \label{eq:sigma_low}
	\frac{d\gamma(\omega)}{d\omega} = \frac{L^2}{2 c_0^2}\left[\left(\frac{\rho_{\mathrm{eff}}(0)}{\rho_0}\right)^2+\left(\frac{K_0}{K_{\mathrm{eff}}(0)}\right)^2\right] +\mathcal{O}(\omega^2),
\end{align}
where we replace $M_{\mathrm{eff}}(\omega) $ by $K_{\mathrm{eff}}(\omega)$ (bulk modulus) for the fluid-based scatterers without shear modes. Because $\left(\frac{\rho_{\mathrm{eff}}(0)}{\rho_0}\right)^2+\left(\frac{K_0}{K_{\mathrm{eff}}(0)}\right)^2\geq 2 {\frac{\rho_{\mathrm{eff}}(0)}{\rho_0}}{\frac{K_0}{K_{\mathrm{eff}}(0)}}$, {\color{black}the coefficient of $\omega^2$ in $\sigma_{\text{ext}}(\omega)$ can be optimally \textit{minimized} as $L^2/c_\mathrm{eff}^2(0)$}, if the static effective and compressibility can be matched
\begin{equation}\label{eq:optimal}
	\frac{\rho_{\mathrm{eff}}(0)}{\rho_0}=\frac{K_0}{K_{\mathrm{eff}}(0)} \;\Leftrightarrow\; Z_{\mathrm{eff}}(0)=Z_0.
\end{equation}
{\color{black}This is a necessary but not sufficient condition for \textit{optimal} scattering. An additional consideration is to ensure that $\frac{\rho_{\mathrm{eff}}(0)}{\rho_0} = \frac{K_0}{K_{\mathrm{eff}}(0)} \neq 1$ ($c_\mathrm{eff}\neq c_0$), so that the static properties contrast with those of the background fluid as a precondition for dynamic modal control. To make this strategy clear,} we construct a Fano resonator, the simplest system coupling a discrete monopole resonance (featured by $K_0/K_{\text{eff}}(\omega)=\frac{\alpha (2c_0/L)}{\omega_m^2 - \omega^2-i \beta \omega}$ ) to a continuum dipole background ($\frac{\rho_\mathrm{eff}(\omega)}{\rho_0}\approx\frac{2c_0}{L}\sqrt{b}$). The different combination of $\rho_\mathrm{eff}(\omega)$ and $K_\mathrm{eff}(\omega)$ can be interpreted as the interference of two oscillators. By taking $k_{\mathrm{eff}} L \ll 1$ in Eq.~(\ref{eq:T}), we can obtain the transmission coefficient for Fano resonance
\begin{equation}\label{eq:fano}
	T(\omega)=\left[  1-i\omega\left(\sqrt{b}+\frac{\alpha}{\omega_m^2-\omega^2-i \beta \omega}\right)\right]  ^{-1}.
\end{equation}
In fact, Ref.~\cite{goffaux2002evidence} derived an exactly same formula of Eq.~(\ref{eq:fano}) via a different approach.  According to Eq.~(\ref{eq:optimal}), to minimize the low-frequency $\omega^2$ coefficient of $\sigma_\mathrm{ext}$, it yields $\sqrt{b}=\alpha/\omega_m^2$ (optimal Fano scattering condition), which results in a \textit{maximum} degree of asymmetry in the transmission lineshape [refer to Ref.~\cite{labelSM}, Sec.~S4].

\begin{figure}[t!]
	\centering
	\includegraphics[width=0.5\textwidth]{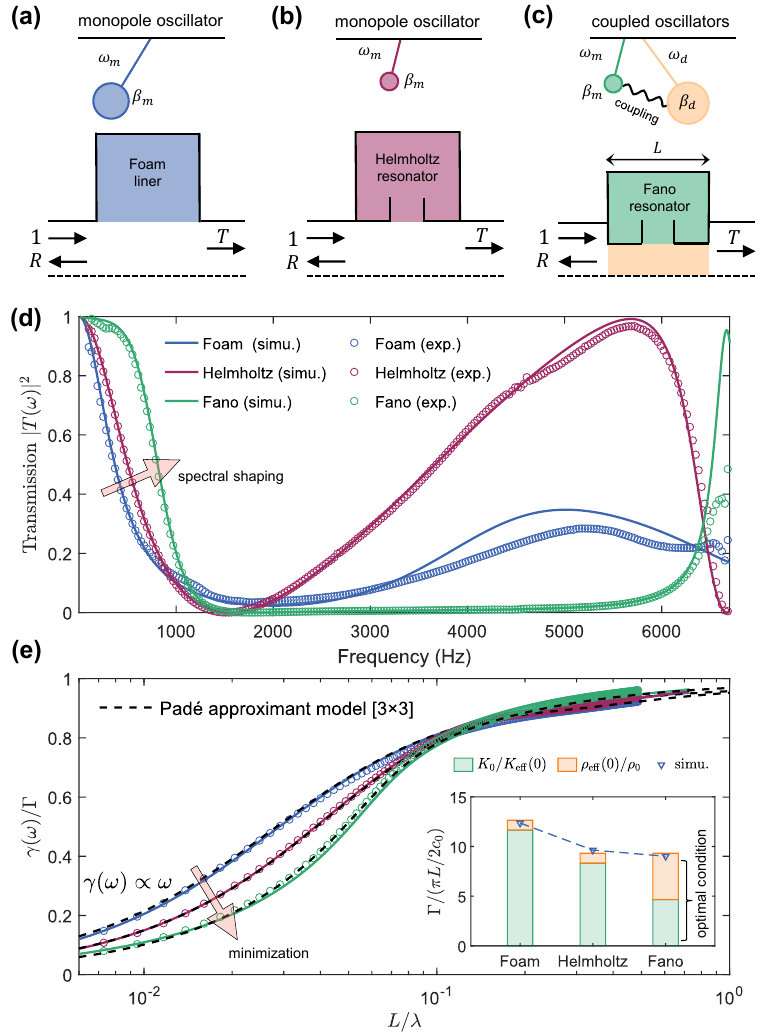}
	\caption{\label{fig:3} \color{black} The experimental validation via airborne sound resonators in ducts. (a) Foam liner (monopole type). (b) Helmholtz resonator (monopole type). (c) Fano resonator (coupled monopole-dipole type). The size of the oscillator sphere represents the amount of dissipation, and the length of the line represents the resonant frequency. Dashed lines represent 2D rotational symmetry axes, and solid lines represent hard boundaries. (d) The simulated (solid lines) and measured (circles) transmission spectra. (e) The cumulative distribution function $\gamma(\omega)$. The inset shows the calculated values of $K_0/K_\mathrm{eff}(0)$ and $\rho_\mathrm{eff}(0)/\rho_0$, along with their ratios and the simulation-extracted values. The experimental $\gamma(\omega)$ below 100 Hz is complemented by low-frequency asymptotic forms derived from simulations (see Ref.~\cite{labelSM} Sec. S4 for details).}
\end{figure}

\textit{Experimental verification.}— To validate the sum rule and its design principle, we fabricated three acoustic resonators within a two-port air duct: a foam liner, a Helmholtz resonator, and a Fano resonator [Figs.~(\ref{fig:3}a–c)]. A useful principle \cite{qu2025duality} to determine their effective properties is
\begin{equation}\label{eq:eff}
	\left\{
	\begin{array}{l}
		{\rho_{\mathrm{eff}}(0)}/{\rho_0} = {1}/{\varphi} + {\Delta L}/{L}\\
		{K_0}/{K_{\mathrm{eff}}(0)} = {V_\mathrm{eff}}/{(S_0 L)}
	\end{array}
	\right.,
\end{equation}
where $\varphi$ is the perforation ratio of the narrowed channel (with $\Delta L$ accounting for the near-field end correction \cite{qu2024analytical}), and $V_\mathrm{eff}$ is the total effective volume of the scatterer, defined with Wood's formula \cite{milton2022theory,ge2025causal}. In Table~(\ref{tab:resonator_models}) of End Matter, we present our simplified model of the three designed resonators with the following features
\begin{enumerate}
	\item Foam liner: $K_0/K_\mathrm{eff}(0)= 11.65$ (due to the volume of additional porous material installed on the sidewalls of the pipe), and $\rho_\mathrm{eff}(0)/\rho_0 \approx 1$ (fully ventilated with $\varphi\approx1$). The foam liner can be treated as a lossy monopole resonator. Porous material region was modelled with Johnson-Champoux-Allard model \cite{yang2017sound}.
	\item Helmholtz resonator: $K_0/K_\mathrm{eff}(0)=8.32$ (with the same air volume as the foam liner, but lower effective compressibility due to the dominant adiabatic process at low frequencies~\cite{ge2025causal}), and $\rho_\mathrm{eff}(0)/\rho_0 \approx 1$. The Helmholtz resonator can be approximated as a lossless monopole resonator.
	\item Fano resonator (the simplest system coupling a discrete monopole resonance to a continuum dipole background): accordingly, $K_0/K_\mathrm{eff}(0)=\rho_\mathrm{eff}(0)/\rho_0 = 4.66$ [$\varphi=25\%$, see the inset of Fig.~(\ref{fig:3}e)]. The narrowed channel results in $\rho_\mathrm{eff}(0)/\rho_0>1$. The air inside the perforated pore supports radiative piston motion (dipole), while the Helmholtz cavity primarily contributes to a monopole resonant mode. Moreover, $\Gamma$ of the Fano resonator equals that of the Helmholtz resonator. 
\end{enumerate}

By using simulation and experiments (details in Ref.~\cite{labelSM}, Sec.~S3 and S4), in Fig.~(\ref{fig:3}d), we confirmed superior sound insulation performance of Fano resonator with optimal scattering condition [Eq.~(\ref{eq:optimal})], with an average measured $\langle\mathrm{TL}\rangle = 21.3\,\mathrm{dB}$ in the frequency range of $1098\,\mathrm{Hz}$–$6174\,\mathrm{Hz}$. However, for Helmholtz resonator, $\langle\mathrm{TL}\rangle = 18.6\,\mathrm{dB}$ in the range $960\,\mathrm{Hz}$–$2332\,\mathrm{Hz}$, and for the foam liner, $\langle\mathrm{TL}\rangle = 13.8\,\mathrm{dB}$ in the range $960\,\mathrm{Hz}$–$3156\,\mathrm{Hz}$. The Fano resonator's total $V_\mathrm{eff}$ is less than that of other two cases, but the TL performance is better due to the design guided by sum rule.

In Fig.~(\ref{fig:3}e), we confirmed that the optimally interfered Fano resonator squeezes more $\sigma_\mathrm{ext}$ resources to a higher frequency band of interest, as characterized by the lowest $\gamma(\omega)$ curve, resulting in the exceptional TL bandwidth shown in Fig.~(\ref{fig:3}d). {\color{black}We emphasize that the sum rule is a universal constraint on the relative bandwidth, because Eq.~(\ref{eq:sum_rule}) can alternatively be written in a dimensionless form: $\int_0^{\infty} \sigma_\mathrm{ext}(\lambda) \, d\left(\frac{\lambda}{L} \right) = \pi^2 \left(\frac{K_0}{M_{\mathrm{eff}}(0)} + \frac{\rho_{\mathrm{eff}}(0)}{\rho_0}\right)$.
If $|T| \to 0$ (high TL) over the operating wavelength range $\lambda \in [\lambda_1, \lambda_2]$, then $\int_{\lambda_1/L}^{\lambda_2/L} \sigma_\mathrm{ext}(\lambda) \, d\left(\frac{\lambda}{L} \right) \approx 2 \frac{\lambda_2 - \lambda_1}{L}$, which reveals how $\sigma_\mathrm{ext}$ is related to the TL bandwidth.} For all cases, with broadband enough spectra data, the measured ultimate bound verified the correctness of acoustic Baldin sum rule: $\gamma(2\pi\times 6500 \mathrm{Hz})/\Gamma$ approaches unity, as shown in Fig.~(\ref{fig:3}e). We further showed that the measured $T(\omega)$ with both phase and amplitude, shown in Fig.~(\ref{fig:4}) of End Matter, follows the Kramers-Kronig relations as well.

Equation (\ref{eq:T_low}) is only an asymptotic form for low frequencies. {\color{black}To overcome the implicit nature of the KK relations and establish an explicit family of causal scattering functions over a broad band, we introduce a Padé approximant model \cite{friedland2012control} of order \([m \times n]\):}
\begin{equation} 
	2\left[1-T(\omega) \right]  = \frac{a_1(-i\omega)+a_2(-i\omega)^2+...+a_m(-i\omega)^m}{1+b_1(-i\omega)+b_2(-i\omega)^2+...+b_n(-i\omega)^n}, \label{eq:Pade}
\end{equation}
where $|T(\omega)|\leq1$ with $m\leq n$ (passivity) and causality prohibits the presence of poles in the upper half-plane $(\operatorname{Im}(\omega)\geq0)$, under the constraint of Routh–Hurwitz criterion \cite{friedland2012control}. If $\omega\to0$ (Taylor series), $2[1-T(\omega)]=\left(-i a_1\right) \omega+\left(a_1 b_1-a_2\right) \omega^2+i\mathcal{O}(\omega^3)$, so, $ a_1 = 2\Gamma/\pi$ (fixed by sum rule's bound). Except $a_1$, other coefficients (real-valued) contribute to the full degree of freedom of spectral shaping. We fitted the three cases in Fig.~(\ref{fig:3}) with a Padé model of minimal order ($m=3$, $n=3$) to capture the low-frequency feature of Fano resonance as per Eq.~(\ref{eq:fano}) [see the fitting details in Ref.~\cite{labelSM}, Sec.~S4]. As shown in Fig.~(\ref{fig:3}e),  $\gamma(\omega)$ from Padé models (dashed lines) also {\color{black}validated the spectral shaping effect}.

\textit{Discussion.}— Usually, to achieve ultra-broadband wave manipulation, multiple resonators must be integrated, thus introducing undesired algorithmic complexity and fabrication challenges. Here, counter-intuitively, our Fano structure is remarkably simple, representing the minimal physical elements required to achieve causally optimal scattering. A sum-rule-based Figure of Merit further confirms that this design ranks among the highest for broadband ventilated silencers \cite{wang2023meta,nguyen2020broadband,jia2025acoustic,lu2025ultra,mei2025reconfigurable,dong2021ultrabroadband,xu2024broadband,fu2025ultrabroadband,jimenez2017rainbow} [see Fig.~(\ref{fig:5}) in End Matter]. {\color{black}Although many theoretical versions of sum rules have been proposed for classical acoustic \cite{meng2022fundamental,norris2018integral} and electromagnetic \cite{padilla2024fundamental,gustafsson2010sum} waves, we reveal the missing link to the Kramer-Kronig relations and Baldin sum rule in quantum field theory, and achieve its direct experimental verification and first application in acoustics.} Although we are only discussing the 1D plane wave scatterings, the relevant theoretical framework can be easily extended to 2D and 3D systems \cite{ramachandran2023bandwidth,martin2024acoustic}.

Beyond acoustics, this work bridges a conceptual gap, revealing the classical counterpart of a fundamental quantum scattering constraint. The underlying framework, described by the family of causal Padé approximants described by Eq.~(\ref{eq:Pade}), is universal. It provides a practical design strategy applicable not only to suppressing waves (as in absorbers \cite{qu2024analytical,qu2025duality,yang2025acoustic}, super-scatterers \cite{ramachandran2023bandwidth}) but to enhancing wave transport (e.g., cloaking \cite{monticone2013cloaked}, impedance matching layers \cite{dong2025soft,dong2020bioinspired}), and even to spectrum customization applications \cite{wang2023meta,qu2023reverberation}.

This work was supported by Jockey Club Trust STEM Lab of Scalable and Sustainable Photonic Manufacturing (GSP181). S. Q thank Seed Fund for Basic Research for New Staff from HKU-URC (No.~103035008). N. X. F. acknowledge the financial support from RGC Strategic Topics Grant (STG3/E-704/23-N), ITC-ITF project (ITP/064/23AP) and startup funding from MILES in HKU-SIRI and the State Key Laboratory of Optical Quantum Materials.

\newpage

\setcounter{equation}{0}
\renewcommand{\theequation}{A\arabic{equation}}
\renewcommand{\thetable}{A\arabic{table}}
\onecolumn
\newpage
\twocolumn

\section*{End Matter}

\begin{table*}[t!] 
	\small
	\captionsetup{width=0.93\textwidth}
	\caption{\label{tab:baldin_comparison}Comparison of Quantum and Acoustic Baldin Sum Rules. Note: for quantum scattering \cite{schumacher2005polarizability} (natural unit with $\hbar=c=1$), $\alpha_{\text{em}}$ denotes the fine-structure constant, and $M,\,Z$ represent the total nucleon mass and charge number, respectively. The mean-square charge radius $r_E^2$ and the effective diamagnetic radius $r_{\text{dia}}^2$ are defined in terms of the constituent parameters: $r_E^2 = \sum_{i} q_i \rho_i^2, \quad r_{\text{dia}}^2 = \sum_{i} \frac{M}{3m_i} q_i^2 \rho_i^2 + \mathbf{D}^2,$ where $q_i$, $m_i$ and $\rho_i$ are the charge fraction, the mass and  the internal coordinate of the constituents inside the hadron. $\mathbf{D}$ represents the electric dipole moment operator, while $D_z$ and $M_z$ denote the $z$-components of the electric and magnetic dipole moments, respectively. The summation over $|n\rangle$ runs through the set of excited states.}
	\centering
	\begin{tabular}{lll}
		\hline\hline
		Physical Context & Quantum Baldin Sum Rule & Acoustic analogy \\ 
		\hline
		Scatterer & Nucleons (Protons / Neutrons) & Meta-atom (Periodic or in-duct) \\
		Eigenvalue & $E_n$ (Energy levels) & $\omega_{m,k}\,,\omega_{d,j}$ (Resonances)\\
		Excitation wave & Photons (Electromagnetic waves) & Sound waves \\ 
		Mathematical formulation & $\displaystyle \int_{\nu_0}^{\infty} \sigma_{\text{ext}}(\nu) \frac{d\nu}{\nu^2} = 2\pi^2(\alpha_e + \alpha_m)$ & $\displaystyle \int_0^{\infty} \sigma_\mathrm{ext}(\omega) \frac{d \omega}{\omega^2} = \Gamma_m + \Gamma_d$ \\
		Optical theorem & $\displaystyle \sigma_{\text{ext}} = \frac{4\pi}{k_0}\text{Im}[f(\theta=0)]$ & $\displaystyle \sigma_{\text{ext}} = 2\text{Re}[1-T(\omega)]$ \\
		Forward scattering  &  $f(\theta=0)$ (3D, far-field spherical wave) &  $T(\omega)-1$ (1D, plane wave) \\
		\multirow{2}{*}{\shortstack[l]{Static parameters\\(electric/magnetic polarizability)\\(monopole/dipole bound term)}} 
		& $\displaystyle\alpha_e=2\alpha_{\text{em}}\sum\limits_{n\not=0}\frac{|\langle n|D_z|0\rangle|^2}{E_n-E_0}+Z^2\alpha_{\text{em}}\frac{\langle0|r_E^2|0\rangle}{3M}$ 
		& $\displaystyle \Gamma_m=\left( \frac{\pi L}{2 c_0}\right) \frac{K_0}{M_{\mathrm{eff}}(0)}=\pi\sum_{k}\frac{\alpha_{m,k}}{\omega_{m,k}^2}$ \\
		& $\displaystyle\alpha_m=2\alpha_{\text{em}}\sum\limits_{n\not=0}\frac{|\langle n|M_z|0\rangle|^2}{E_n-E_0}-\alpha_{\text{em}}\frac{\langle0|r_{\text{dia}}^2|0\rangle}{2M}$ 
		& $\displaystyle\Gamma_d=\left( \frac{\pi L}{2 c_0}\right) \frac{ \rho_{\mathrm{eff}}(0)}{\rho_0}=\pi\sum_{j}\frac{\alpha_{d,j}}{\omega_{d,j}^2}$ \\
		\hline\hline
	\end{tabular}
\end{table*}

\begin{table*}[t!]
	\small
	\captionsetup{width=0.93\textwidth}
	\caption{Effective parameters and transmission  Padé models for different resonator types. For foam liner ($\beta\neq0$) and Helmholtz resonator ($\beta\approx0$), we adopted Padé model $[1\times2] $ with the assumptions $\omega_m^2\ll2 \alpha c_0/L$, $\beta\ll 2c_0/L$ and $\omega_m \ll 2c_0/L$, where $\omega_m$ is the subwavelength monopole resonance frequency. For Fano resonator,   Padé model $[3\times3] $ was utilized [equivalent to Eq.~(\ref{eq:fano})]. General Padé model as per Eq.~(\ref{eq:Pade}) can fit the spectrum of physical realizable scatterers. The theoretical foundation of displayed equations is discussed in Ref.~\cite{labelSM}, Sec.~S2.}
	\centering
	\begin{tabular}{cccc}
		\hline\hline
		Model & Foam or Helmholtz & Fano & General Padé model\\
		\hline
		$\dfrac{\rho_{\text{eff}}(\omega)}{\rho_0}$ 
		& $\approx 1$ 
		& $\approx\frac{2c_0}{L}\sqrt{b} \, (>1)$ 
		& $\sum_{j} \dfrac{\alpha_{d,j}(2c_0/L)}{\omega_{d,j}^2 - \omega^2 - i\beta_{d,j}\omega}$ \\[1.5ex]
		
		$\dfrac{K_0} {K_{\text{eff}}(\omega)}$ 
		& $\textstyle\dfrac{\alpha (2c_0/L)}{\omega_m^2 - \omega^2 - i\beta\omega}$ 
		& $\textstyle\dfrac{\alpha (2c_0/L)}{\omega_m^2 - \omega^2 - i\beta\omega}$ 
		& $\sum_{k} \dfrac{\alpha_{m,k} (2c_0/L)}{\omega_{m,k}^2 - \omega^2 - i\beta_{m,k}\omega}$ \\[1.5ex]
		
		$2[1 - T(\omega)]$ 
		& $\textstyle
		\dfrac{2\alpha(-i\omega)}{
			\omega_m^2 +(\alpha + \beta)(- i\omega) + (-i\omega)^2}$ 
		
		& $\textstyle
		\dfrac{
			2(\sqrt{b}  \omega_m^2 + \alpha)(-i\omega)
			+ 2  \sqrt{b} \beta(-i\omega)^2
			+ 2  \sqrt{b}(-i\omega)^3
		}{
			\omega_m^2 +( \sqrt{b} \omega_m^2 + \alpha + \beta)(-i\omega)
			+ ( \sqrt{b}  \beta + 1 )(-i\omega)^2 
			+ \sqrt{b} (-i\omega)^3 
		}$ 
		
		& $\textstyle\dfrac{\sum_{j=1}^{m} a_j(-i\omega)^j}{1 + \sum_{k=1}^{n} b_k(-i\omega)^k}$ \\
		
		\hline\hline
	\end{tabular}
	\label{tab:resonator_models}
\end{table*}

\textit{Appendix A. Quantum vs acoustic Baldin sum rule.}— Table~(\ref{tab:baldin_comparison}) elucidates the analogy between the microscopic nucleon scattering parameters and the macroscopic acoustic effective properties. Even though quantum and acoustic scatterers are quite different, spanning micro- and macro-scales, Compton scattering~\cite{schumacher2005polarizability} and general acoustic scattering share the physical picture of wave–matter interaction. Moreover, the causality of scattering leads to the compliance with the Kramers–Kronig relations, and thus to a remarkably similar integral dispersion relation, namely the Baldin sum rule. The energy levels of the nucleons ($E_n$) are analogous to the eigenfrequencies of an acoustic resonator ($\omega_{d,j}$ and $\omega_{m,k}$). The fine-structure constant $\alpha_{em}$ has its counterpart in acoustics as ${\pi}/{2 c_0}$. Due to the symmetry of the atomic states $|n\rangle$, the static polarizability can be categorized into electric and magnetic types. Similarly, the symmetrical and anti-symmetrical pressure fields of an acoustic resonator define the effective modulus and density according to homogenization theory~\cite{yang2014homogenization}.

However, quantum and classical acoustic systems also exhibit some differences. For example, nucleons possess a ground state ($|0\rangle$) with energy $E_0$, which determines the threshold for excitation photon energy $\nu_0$. In contrast, in acoustics, there is usually no such lowest energy level (this makes low-frequency noise mitigation harder). Moreover, the energy of a photon is proportional to $\omega$, whereas for sound plane waves the relation is quadratic ($\propto \omega^2$).

\begin{figure}[b!]
	\centering
	\includegraphics[width=0.5\textwidth]{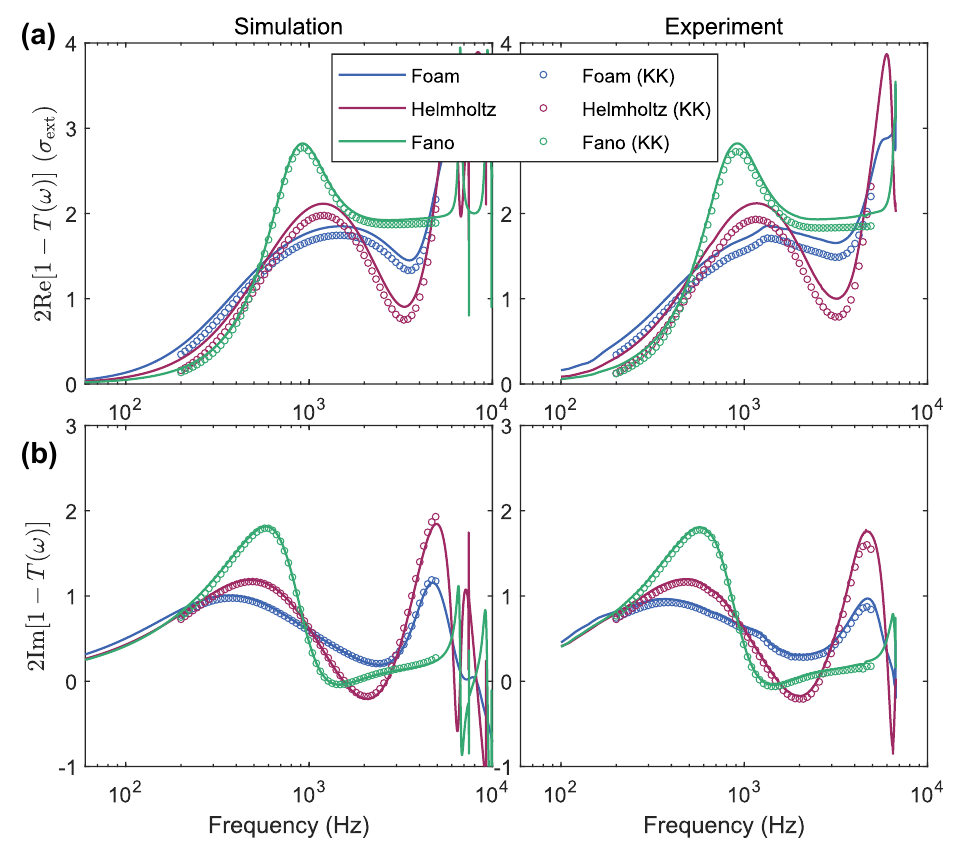}
	\caption{\label{fig:4} \color{black} The direct verification of the Kramers–Kronig (KK) relations using transmission data from FEM simulations (left) and experiments (right). (a) $2\mathrm{Re}[1 - T(\omega)]$ or $\sigma_\mathrm{ext}(\omega)$ and its KK-generated counterpart (circles). (b) $2\mathrm{Im}[1 - T(\omega)]$ and its KK-generated counterpart (circles).}
\end{figure}

\begin{figure}[b!]
	\centering
	\includegraphics[width=0.5\textwidth]{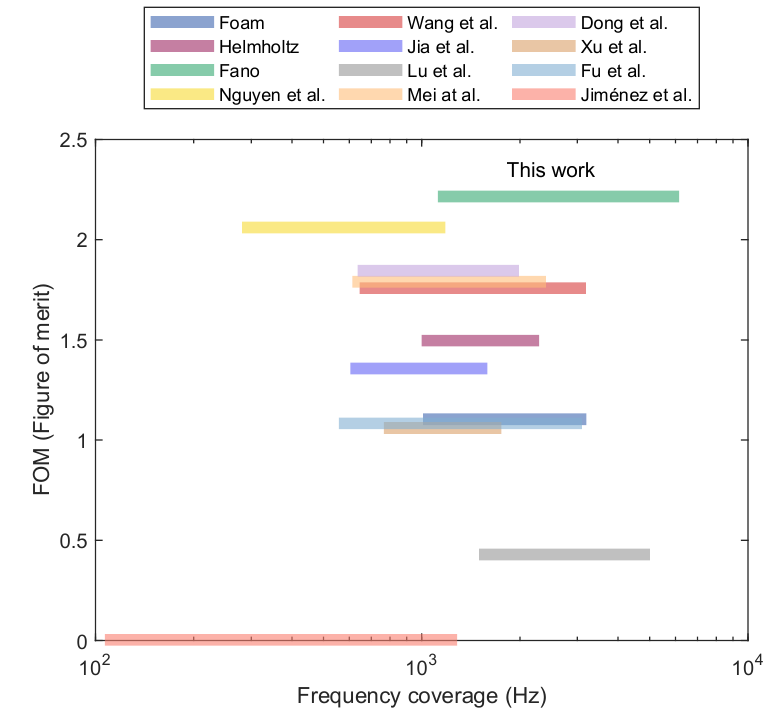}
	\caption{\label{fig:5} \color{black} The Figure of Merit [see Eq.~(\ref{eq:fom})] vs the frequency coverage (beyond 10 dB transmission loss). The data are from our experiments (the sample thickness was $L=2.5\,\mathrm{cm}$ for all three cases) and other reported works \cite{wang2023meta,nguyen2020broadband,jia2025acoustic,lu2025ultra,mei2025reconfigurable,dong2021ultrabroadband,xu2024broadband,fu2025ultrabroadband,jimenez2017rainbow}.}
\end{figure}

\textit{Appendix B. Direct verification of Kramers--Kronig relations.}— The KK relations are evaluated for a set of discrete angular frequencies $\omega_j$ (obtained from simulations or experiments), using the standard discrete summation form:
\begin{subequations}
	\begin{align}
		2\mathrm{Re}[T(\omega_i) - 1] &\approx \frac{2}{\pi} \sum_{j \ne i} \frac{ 2\omega_j\,\mathrm{Im}[T(\omega_j) - 1]}{\omega_j^2 - \omega_i^2}\, \Delta\omega_j, \\
		2\mathrm{Im}[T(\omega_i) - 1] &\approx -\frac{2\omega_i}{\pi} \sum_{j \ne i} \frac{2\,\mathrm{Re}[T(\omega_j) - 1]}{\omega_j^2 - \omega_i^2} \, \Delta\omega_j,
	\end{align}
\end{subequations}
where $\Delta\omega_j = \omega_{j+1} - \omega_j$. {\color{black}Direct numerical integrations were also performed over evaluation points \(\omega_i\) by computing principal-value integrals in two separate intervals, excluding a symmetric exclusion band \(\omega \in \big[(1 - \varepsilon)\omega_i,\ (1 + \varepsilon)\omega_i\big]\) with \(\varepsilon = 10^{-5}\). For simulations, the integration range of \(\omega\) is given by \((a, b) = \left(10 \cdot 2\pi,\ 10^4 \cdot 2\pi\right)\ \mathrm{rad/s}\); for experiments, \((a, b) = \left(100 \cdot 2\pi,\ 6500 \cdot 2\pi\right)\ \mathrm{rad/s}\). The sampling points \(\omega_i\) are defined by a logarithmically spaced grid \(f_i = \exp\big(\log(200): 0.05: \log(5000)\big)\ \mathrm{Hz}\) (65 points), corresponding to \(\omega_i = 2\pi f_i\). This choice balances resolution at both low and high frequencies. The function \(T(\omega)\) was linearly interpolated from the discrete data prior to integration. No smoothing or filtering was applied. Figure~\ref{fig:4} illustrates that the results obtained from the KK relations (circles) agree remarkably well with the original transmission data (solid lines). At low frequencies, the extinction cross-section follows the trend \(\mathrm{foam} > \mathrm{Helmholtz} > \mathrm{Fano}\), consistent with the larger damping \(\Gamma\) in foam. Helmholtz and Fano resonators share the same \(\Gamma\), but only the latter satisfies the static match condition in Eq.~(\ref{eq:optimal}).}

\textit{Appendix C. Figure of merit calculation.}— By replacing $\omega$ with the wavelength $\lambda$ (defined as $2\pi c_0/\omega$), Eq.~(\ref{eq:sum_rule}) can be reformulated as: $\int_0^{\infty} \sigma_\mathrm{ext}(\omega) \, d\lambda = \Gamma^{\prime}$, where $\Gamma^{\prime} = 2\pi c_0 \Gamma$. Using closed-form models as per Eq.~(\ref{eq:eff}), for air-based ventilative resonators [e.g., Fig.~(\ref{fig:3})], this expression can be approximated as $\Gamma^{\prime} \approx \pi^2 L \left(V_\mathrm{eff}/S_0 + L/\varphi \right)$ (end correction omitted). Based on this, to evaluate the performance of our experimental samples and compare them with previously reported works~\cite{wang2023meta,nguyen2020broadband,jia2025acoustic,lu2025ultra,mei2025reconfigurable,dong2021ultrabroadband,xu2024broadband,fu2025ultrabroadband,jimenez2017rainbow}, we introduce a sum-rule-based figure of merit (calculated results are shown in Fig.~\ref{fig:5}):
\begin{equation} \label{eq:fom}
	\mathrm{FOM} = \frac{\langle\mathrm{TL}\rangle \left(\lambda_2 - \lambda_1\right)}{\pi^2 L \left(V_\mathrm{eff}/S_0 + L/\varphi \right)},
\end{equation}
where $\langle\mathrm{TL}\rangle$ denotes the average transmission loss over the sound reduction band (defined as exceeding 10 dB from frequency $f_1$ to $f_2$). In this expression, the wavelengths are given by $\lambda_2 = c_0/f_1$ and $\lambda_1 = c_0/f_2$. Unlike Eq.~(\ref{eq:sum_rule}), Eq.~(\ref{eq:fom}) enables evaluation using limited-band TL data, since $\sigma_\mathrm{ext}(\omega)$, which depends on the phase information of $T(\omega)$, is typically unavailable in reported studies. Our design principle is that the numerator of the FOM quantifies the overall sound insulation performance (taking into account both bandwidth and attenuation level), while the denominator represents the resources consumed by the ventilated metamaterials (thickness $L$, ventilation rate $\varphi$ and volume $V_\mathrm{eff}$).

{\color{black}\textit{Appendix D. Relation to existing integral identities.} Here, we clarify the relationship between our sum rule and the identities derived by Norris~\cite{norris2018integral} and Meng et al.~\cite{meng2022fundamental}. First, regarding the high-frequency limit, Eq.~(33) in Ref.~\cite{norris2018integral} and Eq.~(3.25) in Ref.~\cite{meng2022fundamental} rely on a finite dynamic sound speed $c_\infty$ as $\omega \to \infty$. Defining a precise $c_\infty$ is often challenging in metamaterials due to the breakdown of homogenization. In contrast, our formulation avoids high-frequency dynamic parameters by anchoring the sum rule solely to robust static effective properties [$\rho_{\text{eff}}(0)$ and $M_{\text{eff}}(0)$]. Second, the transmission coefficient $T_{\text{Norris}}$ in Ref.~\cite{norris2018integral} differs from our $T(\omega)$ by a phase factor associated with wave propagation. The relationship is given by $T(\omega) = T_{\text{Norris}} e^{-i\omega \tau'}$, where $\tau' = L/c_{\text{eff}}(0)$ represents the wave travel time through the effective medium. Consequently, while previous bounds appear in terms of polarizabilities, our formulation naturally incorporates the thickness $L$ via this phase relation, thus avoiding the problem noncausal behavior reported in Ref.~\cite{norris2015acoustic}.}
\end{document}